\documentclass[prb,aps,amsmath,amssymb,showpacs,floatfix,twocolumn]{revtex4-1}
\usepackage{graphicx,epsfig}
\usepackage[usenames]{color}
\newcommand{\beq}{\begin{equation}}
\newcommand{\eeq}{\end{equation}}
\newcommand{\bea}{\begin{eqnarray}}
\newcommand{\eea}{\end{eqnarray}}
\newcommand{\mprb}[1]{#1}
\newcommand{\mprbnew}[1]{#1}
\newcommand{\mprbnewnew}[1]{#1}

\begin{document}
\title{Resonant inelastic x-ray scattering as \mprbnewnew{a probe of band structure effects in cuprates}}
\author{M. Kan\'asz-Nagy$^{1}$, Y. Shi$^{2}$, I. Klich$^{2}$ and E. A. Demler$^1$}
\affiliation{$^1$Department of Physics, Harvard University, Cambridge, MA 02138, U.S.A}
\affiliation{$^2$Department of Physics, University of Virginia, Charlottesville, VA 22904}

\begin{abstract}
We analyze within quasi-particle theory a recent resonant inelastic x-ray scattering (RIXS) experiment on $\mathrm{YBa_2Cu_3O_{6+x}}$ with the incoming photon energy detuned at several values from the resonance maximum [Minola {\it et al.}, Phys. Rev. Lett. {\bf 114}, 217003 (2015)]. Surprisingly, the data shows much weaker dependence on detuning than expected from recent measurements on a different cuprate superconductor, $\mathrm{Bi_2Sr_2CuO_{6+x}}$ [Guarise {\it et al.}, Nat. Commun. {\bf 5}, 5760 (2014)]. We demonstrate here, that this
discrepancy, originally attributed to collective magnetic excitations, can be understood in terms of the differences between the band structures of these materials. We find good agreement between theory and experiment over a large range of dopings, both in the underdoped and in the overdoped regime. Moreover, we demonstrate that the RIXS signal depends sensitively on excitations at energies well above the Fermi surface, that are inaccessible to traditionally used band structure probes, such as angle-resolved photemisson spectroscopy. This makes RIXS a powerful probe of band structure, not suffering from surface preparation problems and small sample sizes, making it potentially applicable to \mprb{a number of cuprate materials.}
\end{abstract}

\maketitle

Despite the technological and scientific importance of cuprate high-temperature superconductors, little is known about their overall quasi-particle band structures. Although density functional theory predicts quasi-particle dispersion near the Fermi surface reasonably well, it cannot reliably capture the effect of electron-electron correlations, and thus largely overestimates the excitation energies near the top of the band.~\cite{LDA_calculations}
On the experimental side, traditionally used band structure probes, such as angle-resolved photoemission sectroscopy (ARPES)~\cite{ARPES_review} and quantum oscillation measurements~\cite{Q_oscillation_1, Q_oscillation_2} probe excitations mostly in the vicinity of the Fermi surface, and provide little information on the higher energy part of the band. Moreover, in the case of $\mathrm{YBa_2Cu_3O_{6+x}}$ ($\mathrm{YBCO}_{6+x}$), ARPES suffers from significant surface preparation problems, the so-called polar catastrophe.~\cite{BS_Norman} Thus, current band structure models  of YBCO differ significantly at higher energies.~\cite{BS12,BS3,BS_Norman}
In contrast, resonant inelastic x-ray scattering (RIXS) of transition metal oxides provides a momentum-resolved access to various electronic, magnetic and phononic excitations in a large energy range~\cite{electronic_RIXS,monney,magnetic_RIXS,phononic_RIXS} with an unprecedented sensitivity. Moreover, it also operates on small sample sizes and even films,~\cite{RIXS_film} without suffering from surface quality problems.~\cite{RIXS_review} This provides a unique opportunity to study quasi-particle band structure~\mprbnew{\cite{ Carlisle, Veenendaal, Shirley, Carlisle2, Ahn}} in the optimally doped and overdoped regime, where the dominant excitations are described by electronic quasi-particles.~\cite{electronicQPs,Fermi_surface_HTC1,Fermi_surface_HTC2, Fermi_surface_HTC3, Benjamin}
\mprbnew{In this regime, ARPES-based tight binding models can be tested, by comparing their theoretical RIXS spectra with experimental data.}

\mprb{Despite the recent surge of experiments on doped cuprates, the theoretical description of RIXS in these materials is still debated, and thus, the interpretation of experimental data is often unclear.} Although RIXS measurements have been performed over a wide range of dopings, theoretical work originally focused on the antiferromagnetic part of the cuprate phase diagram,  predicting collective magnetic excitations as the primary contributors to RIXS intensities.~\cite{spin_susc_RIXS_calculations0,spin_susc_RIXS_calculations1,spin_susc_RIXS_calculations2,spin_susc_RIXS_calculations3,
spin_susc_RIXS_calculations4,sr2cuo3_spin_charge}
As a result, experimental data is often interpreted in terms of these models, even in the overdoped regime, where Fermi liquid behavior is expected.~\cite{electronicQPs, Fermi_surface_HTC1,Fermi_surface_HTC2, Fermi_surface_HTC3}
In contrast to these models, recent work by D. Benjamin {\it et al.} modeled the RIXS process in this regime in terms of non-interacting electronic quasi-particles,~\cite{Benjamin} and found good agreement with recent experiments on $\mathrm{Bi_2Sr_2CuO_{6+x}}$ (Bi-2212).~\cite{D_Benjamin_exp1,D_Benjamin_exp2} Although the measured RIXS peaks had originally been attributed to magnetic excitations,
the authors showed that they are well described in terms of band structure physics, combined with an orthogonality catastrophe type many-body effect of the photoexcited core hole on the Fermi sea.~\mprb{\cite{Ahn, Anderson, Nozieres}}
As an experimentally testable prediction of the theory, they also calculated how the  position of the RIXS peaks change as one tunes the incoming photon energy away from the resonance maximum, and found significant shifts in the peak position, as a prediction of band structure theory.
This fluorescent behavior has been confirmed in the experimental study of Bi-2212 by Guarise {\it et al.},~\cite{Ronnow_vdBrink} and similar effects have been found for $\rm Sr_2CuO_2Cl_2$ and $\rm La_2CuO_4$ by Abbamonte {\it et al.}~\cite{Abbamonte}

In order to test these predictions, recent experimental work by Minola {\it et al.}~\cite{Minola} investigated the dependence of the RIXS intensity
on incoming photon energy \mprb{at the Cu-$L_3$ edge of} a different material, $\mathrm{YBCO}_{6+x}$. However, they found the shifts of the 
peaks to be insignificant, thus interpreted their results in terms of collective paramagnon excitations, \mprb{and claimed to falsify the quasi-particle approach of Ref.~\onlinecite{Benjamin}}. In this work, we resolve this puzzle by pointing out that the small shift of the peaks is simply
explained by the differences between YBCO and Bi-2212 band structures. We show that the experimental results are well described by band structure physics for a wide range of dopings \mprb{($x\gtrsim 0.79$), from the slightly underdoped to the overdoped side}.
Moreover, comparing our results to experimental data allows us to test different tight-binding models of YBCO.~\cite{BS3,BS12,BS_Norman}
These models were obtained as fits to ARPES measurements of a small energy window near the Fermi surface. 
It is thus not surprising, that they exhibit factor of three differences in energy near the top of the band. 
\mprbnew{ By comparing the RIXS signatures of the different band structure models, we were able to choose the band structure that gives the best agreement with the experimental data of Ref.~\onlinecite{Minola}. This tight binding model thus provides the most accurate description of the high energy excitations of YBCO during the RIXS processes of the experiment.}

In order to determine the RIXS response, we use a tight-binding Hamiltonian $H = \sum_{j l \, \sigma} \, t_{jl} \, d^\dagger_{j \sigma} d_{l \sigma}$, with $d_{j \sigma}$  annihilating a conduction electron of spin $\sigma$ at site $j$. The hopping amplitudes $t_{jl}$ for YBCO are given by $\left( t_{10}, \, t_{11}, \, t_{20}, \, t_{21} \right) = \left(-105, \, 29, \, -25, \, 4  \right) {\rm meV}$.~\cite{BS3}
The RIXS process at the Cu-$L_3$ edge starts with a photoexcitation of a $2p$ core electron into a $3d_{x^2-y^2}$ state in the conduction band, leaving a positively charged core hole behind. This hole is then refilled again with an electron, and a photon is emitted, with energy loss $\Delta \omega$ and momentum transfer $\bf q$ with respect to the incoming photon.~\cite{Benjamin}
The scattering process is given by
\beq
I\left(\Delta\omega,\, \omega, \, {\bf q}_\parallel\right) \, \propto \, \sum_{f} \left| T_{i \to f} \right|^2 \, \delta(E_f - E_i -\Delta\omega),
\label{eq:RIXS_intensity}
\eeq
where $\omega$ denotes the incoming photon energy, ${\bf q}_\parallel$ refers to the momentum loss parallel to the sample surface,
and $E_{i(f)}$ stands for the energy of the initial (final) many-body state $\left| i \right\rangle$ $\left( \left| f \right\rangle \right)$.
In real experiments, the Dirac delta function is broadened due to the finite energy resolution of the measurement apparatus.
The transition amplitudes are given by the Kramers-Heisenberg  formula~\cite{spin_susc_RIXS_calculations1, spin_susc_RIXS_calculations2}
\beq
T_{i \to f} = \sum_{j\sigma \sigma^\prime} \chi_{\sigma\sigma^\prime} \, e^{i {\bf q}_\parallel {\bf r}_j} 
\langle f | d_{j\sigma} \left( H_j - E_i - \omega + i\Gamma \right)^{-1} d_{j \sigma^\prime}^\dagger | i \rangle,
\nonumber
\eeq
where the prefactors $\chi_{\sigma\sigma^\prime}$ originate from matrix elements of the optical transitions, and therefore depend strongly on the scattering geometry, as well as on photon polarizations.~\cite{Benjamin} The Hamiltonian $H_j = H + V_j$ contains the  $V_j =  - U_0 \sum_\sigma d_{j\sigma}^\dagger d_{j\sigma}$ potential of the positively charged core hole  at site $j$, and $\Gamma$ denotes the inverse lifetime of the hole, usually on the order of $\Gamma \sim$ 250-500 meV for cuprates.~\cite{Benjamin, QPlifetime_footnote}
In order to determine the RIXS response numerically, we follow the lines of  Ref.~\onlinecite{Benjamin}. We sum over all final states, and rewrite the RIXS intensity in terms of time integrals
\bea
I \,& \propto & \, \int_{-\infty}^\infty ds \, \int_{0}^\infty dt \; \int_{0}^\infty d\tau \,
e^{(-i\omega - \Gamma)t} \, e^{(i\omega - \Gamma)\tau} \, e^{-i \Delta\omega \, s} 
\nonumber \\
&\times & \, \sum_{j l \sigma \sigma^\prime {\tilde \sigma} {\tilde \sigma}^\prime}e^{-i {\bf q}_\parallel ({\bf r}_j  - {\bf r}_l)}
\chi_{\sigma\sigma^\prime}^* \chi_{{\tilde \sigma}{\tilde \sigma}^\prime}
\; S^{jl}_{\sigma \sigma^\prime {\tilde \sigma} {\tilde \sigma}^\prime} (s, t, \tau)
\eea
with the four point correlation function
\beq
S^{jl}_{\sigma \sigma^\prime {\tilde \sigma} {\tilde \sigma}^\prime} = 
\left\langle
d_{j\sigma} e^{-i H_j \tau} d_{j\sigma^\prime}^\dagger \,  e^{i H s} \,  d_{l{\tilde \sigma}} 
e^{i H_l t}  d_{l{\tilde \sigma}^\prime}^\dagger \, e^{i H (\tau - t -s)}
\right\rangle,
\eeq
where the expectation values are taken with respect to the Fermi sea, $\left| i \right\rangle = \left| FS \right\rangle$.
Importantly, the presence of the core hole perturbs the entire Fermi sea, creating infinitely many particle-hole excitations, during
the RIXS process. Eq.~\eqref{eq:RIXS_intensity} therefore contains significant many-body contributions, that are related to the 
orthogonality catastrophe problem, originally investigated in the context of the x-ray edge singularity of 
metals.~\cite{OC_physics1, OC_physics2, Nozieres}
Making use of the determinant formulas of Ref.~\onlinecite{Klich_formulas},
the non-trivial many-body correlator $S^{jl}_{\sigma \sigma^\prime {\tilde \sigma} {\tilde \sigma}^\prime}$ can be rewritten exactly 
in terms of single particle Hamiltonian matrices $\left(h_j\right)_{l m} = t_{l m} - U_0 \, \delta_{jl} \, \delta_{l m}$.
This single particle form vastly simplifies our numerical computations (see Ref.~\onlinecite{Benjamin} for details).

\begin{figure}[t]
\includegraphics[width=8.5cm,clip=true]{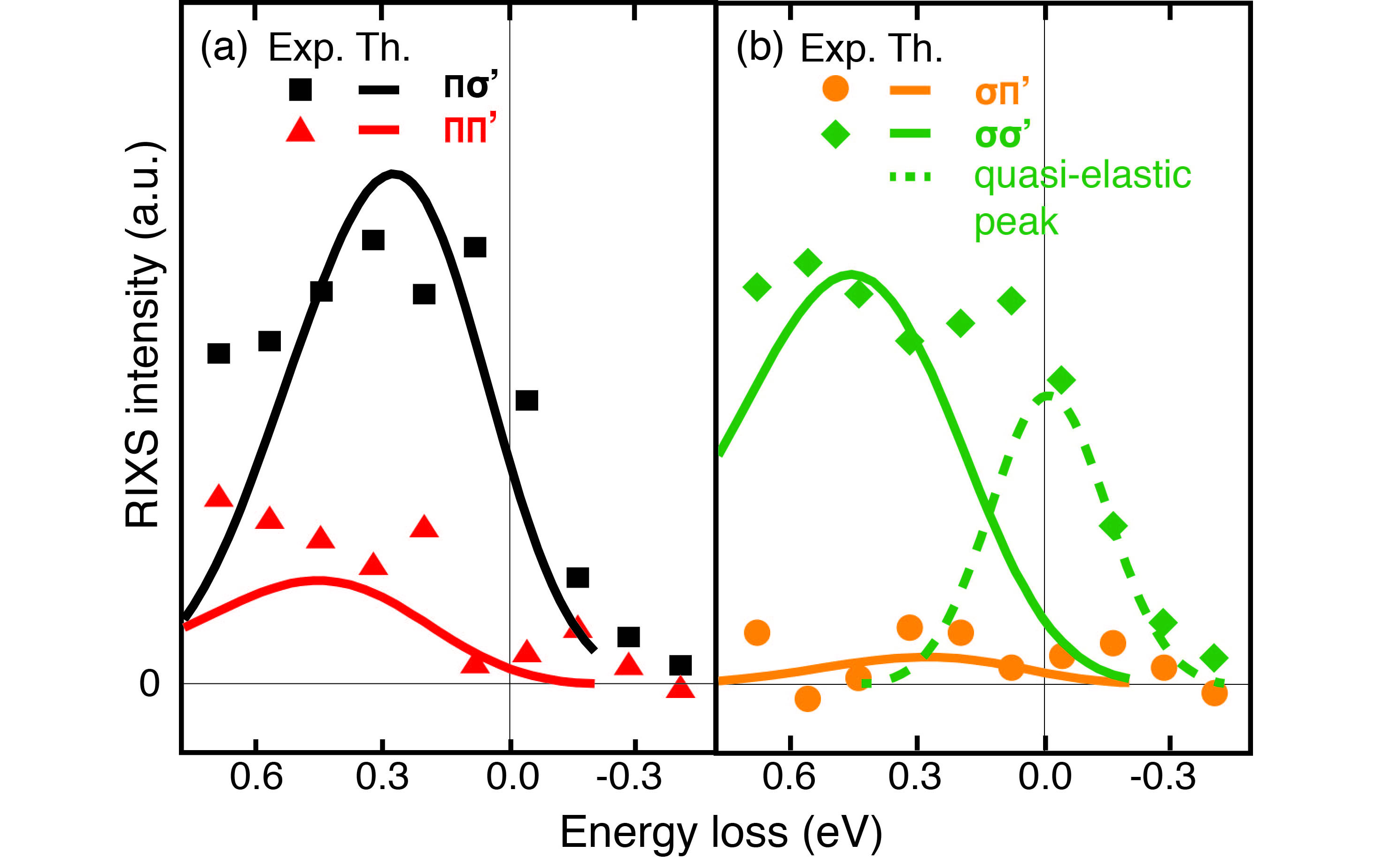}
\caption{(Color online.) 
Polarization resolved RIXS spectra with incoming $\pi$ (a) and $\sigma$ (b) polarization
on overdoped ${\rm YBCO+Ca}$ at the Cu-$L_3$ edge. Symbols (full lines) denote   
experimental (theoretical) data, whereas the dashed line in (b) corresponds to Gaussian fit 
to the quasi-elastic part of the $\sigma\sigma^\prime$ channel. 
RIXS intensity is dominated by SF/NSF processes in the scattering geometry of
(a)/(b), respectively.
[Parameters: ${\bf q}_\parallel = 2\pi \,(0.37, 0)$, $U_0 = 1 \, {\rm eV}$, 
$\Gamma = 250 \, {\rm meV}$, energy resolution 95 meV HWHM, lattice size $22 \times 22$.]}
\label{fig:polarization_resolved}
\end{figure}

With the aid of the polarization matrix $ \boldsymbol \chi$, the RIXS intensity can be decomposed into
non-spin-flip (NSF) and spin-flip (SF) contributions, 
\beq
I = \left| \chi_{\rm NSF} \right|^2 \, I_{\rm NSF} + \left| \chi_{\rm SF} \right|^2 \, I_{\rm SF},
\eeq
where the weights of the NSF and SF channels are given by the diagonal and off-diagonal elements of the polarization matrix, respectively.~\cite{Benjamin}
These channels can be decomposed in an experiment by polarization analysis of the incoming and outgoing photons.
Since a single spin-flip is accompanied by a $90 \, ^{\circ}$ rotation of the photon polarization,
the spin flip channel always corresponds to the $\pi \sigma^\prime$ or $\sigma \pi^\prime$ scattering,
where $\pi$ ($\pi^\prime$) and $\sigma$ ($\sigma^\prime$) denote the incoming (outgoing) 
polarizations.
In contrast, non-spin-flip processes exclusively contribute to the $\sigma\sigma^\prime$ and $\pi \pi^\prime$
channels.~\cite{Minola}
The NSF and SF channels often exhibit significantly different peak structures, and the latter is thus often
associated with spin excitations. However, the difference between the NSF and SF contributions can 
be explained within band structure theory, as originating from the spin selective screening of the core 
hole by the photoexcited electron, as was shown in Ref.~\onlinecite{Benjamin}.

\begin{figure}[t]
\includegraphics[width=8.5cm,clip=true]{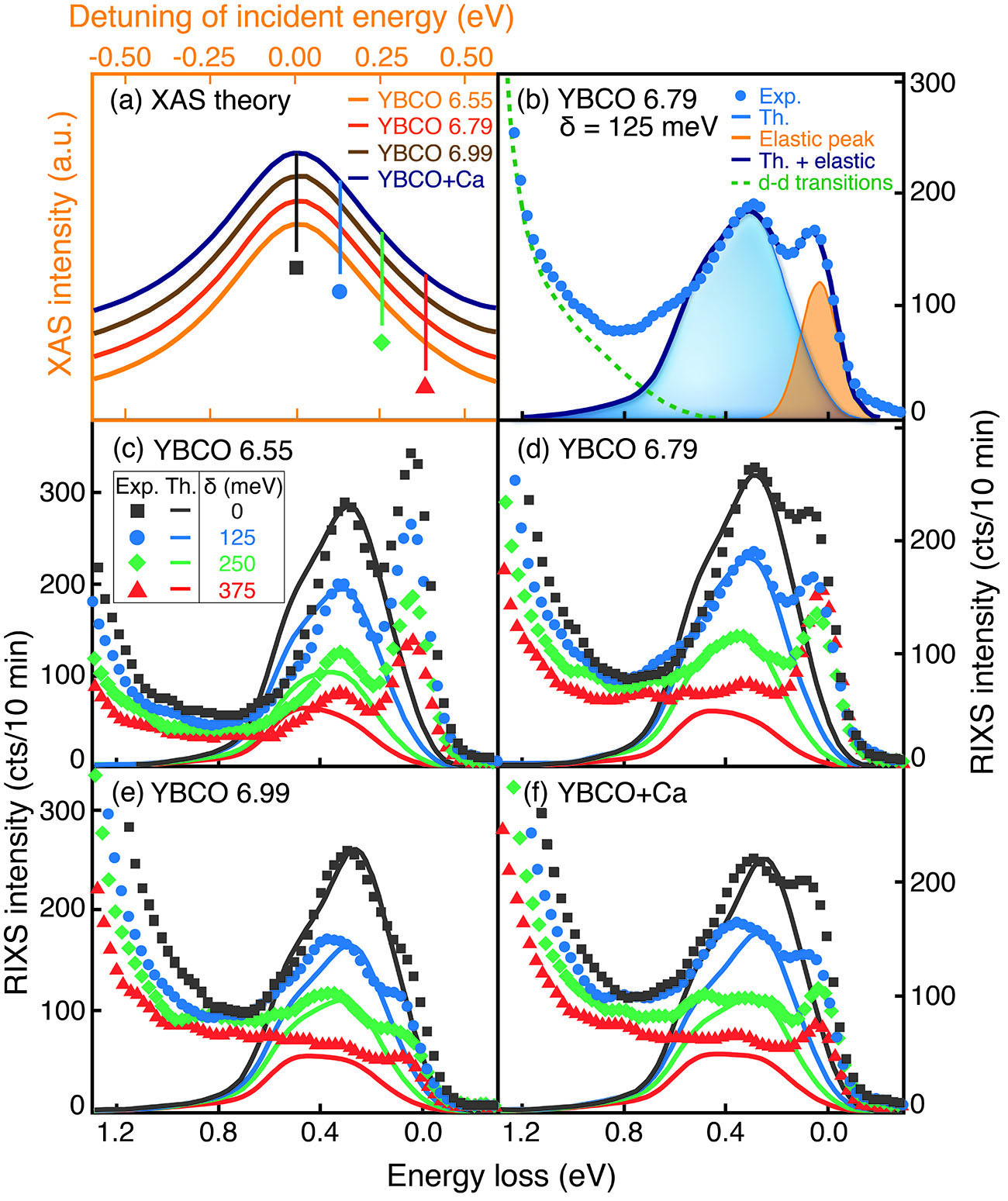}
\caption{(Color online.) 
RIXS spectra of YBCO measured with $\pi$ incoming polarization. The incoming photon energy is
detuned $\delta = (0, 125, 250, 375)$ meV away from the XAS maximum, shown in (a).
(b) The experimental RIXS data shows pronounced quasi-elastic peaks together with another peak near 300 meV,
fit by our theoretical model, as well as a high-energy tail of $dd$ excitaitons, as indicated schematically by the dashed green line.
(c-e) RIXS spectra at different dopings and detunings, with experimental (theoretical) data denoted by symbols (full lines). 
\mprbnew{Elastic contributions and d-d excitations are not taken into account in the theoretical curves.}
[Parameters and scattering geometry are identical to the ones in Fig.~\ref{fig:polarization_resolved}~(a), 
however, outgoing polarizations are mixed.]
}
\label{fig:SF_1D}
\end{figure}

Fig.~\ref{fig:polarization_resolved} compares our theoretical calculations to the polarization resolved experimental data of Ref.~\onlinecite{Minola}
for a highly overdoped doped ${\rm YBCO + Ca}$ sample (${\rm Y_{0.85}Ca_{0.15}Ba_2Cu_3O}_{6+x}$, doping level $p\sim 0.21$).
The spectra were taken at the resonance of the x-ray absorption spectrum approximated as,~\cite{RIXS_review,spin_susc_RIXS_calculations1}
\beq
{\rm XAS} (\omega) \, = \, \int_{-\infty}^\infty d{\Delta\omega} \, I\left(\Delta\omega,\, \omega, \, {\bf q} = 0\right),
\eeq
at the Cu-$L_3$ edge. This, and all further measurements were taken at a momentum transfer ${\bf q}_\parallel = 2\pi \, (0.37, 0) $. In order to investigate non-spin-flip (NSF) and spin-flip (SF) channels separately, the geometries of the incoming and scattered photons were chosen such that the RIXS signal is dominated by the NSF (SF) channel, with $|\chi_{\rm SF}|^2 / (|\chi_{\rm NSF}|^2 + |\chi_{\rm SF}|^2) = 3 \%$ ($68 \%$) for the right (left) figure.
This led to a more pronounced separation of the NSF and SF channels than in Refs.~\onlinecite{Ronnow_vdBrink,Abbamonte}. Besides the quasi-elastic peak near zero energy loss in the $\sigma \sigma^\prime$ (NSF) channel, \mprb{corresponding to phonons and sample imperfections},~\cite{RIXS_review} we see pronounced inelastic peaks near 550 (400) meV in the NSF (SF) channel. Although Ref.~\onlinecite{Minola} interpreted the inelastic contributions in the SF channels as originating from collective magnetic modes, we find that the spectrum can be well described within our band structure model. \mprbnew{As both the experimental data and the theoretical curves show, the NSF intensity is shifted to higher energy losses than the peak in the SF channel. This effect arises from the spin-selective screening of the core hole within the quasi-particle model, as was explained in Ref.~\onlinecite{Benjamin}.}

Ref.~\onlinecite{Minola} also performed a detailed RIXS study of several ${\rm YBCO}_{6+x}$ samples 
from the underdoped to the overdoped regime, with $\pi$ incoming and mixed outgoing polarization.
Using the same geometry as in Fig.~\ref{fig:polarization_resolved}~(b), led to dominantly SF scattering.
In order to investigate the effect of the incoming photon energy on the RIXS signal, 
$\omega$ was tuned $(0, 125, 250, 375) \, {\rm meV}$ away from the XAS maximum, shown in Fig.~\ref{fig:SF_1D}~(a).
Similarly to Fig.~\ref{fig:polarization_resolved}~(b), we find a quasi-elastic peak near zero energy, 
and a secondary peak near $350\, {\rm meV}$, as well as a tail of high-energy $dd$ transitions,~\cite{RIXS_review} not taken into account by our model, see Fig.~\ref{fig:SF_1D}~(b).
Up to an overall normalization factor, theoretical spectra (full lines) fit the experimental data reliably,
as shown in \mprb{Figs.~\ref{fig:SF_1D}~(d-f)}.
Both the peak positions and the widths are reproduced for a large range of detunings: underdoped 
\mprb{${\rm YBCO_{6.79}} \, (p \sim 0.142)$,} optimally doped ${\rm YBCO_{6.99}} \, (p \sim 0.189)$ and overdoped
and ${\rm YBCO + Ca} \, (p \sim 0.21)$.
\mprb{On the other side of the phase diagram, where the quasi-particle description is no longer valid, our model fails to reproduce the sharp RIXS signal found in the antiferromagnetic sample ${\rm YBCO}_{6.10}$, which expected to arise from collective magnetic excitations (see Appendix~\ref{app:AF_phase}). Furthermore, the quasi-particle model overestimates the peak widths in the strongly underdoped sample ${\rm YBCO_{6.55}} \, (p \sim 0.114)$ either (see Fig.~\ref{fig:SF_1D}~(c)). This agrees with our expectations, that quasi-particle theory should be most reliable on the overdoped side, and should not be applicable in the strongly underdoped phase, due to strong interactions between electronic and magnetic excitations.~\cite{Devereaux, Keimer_more_friendly}}

\begin{figure}[t]
\includegraphics[width=8.5cm,clip=true]{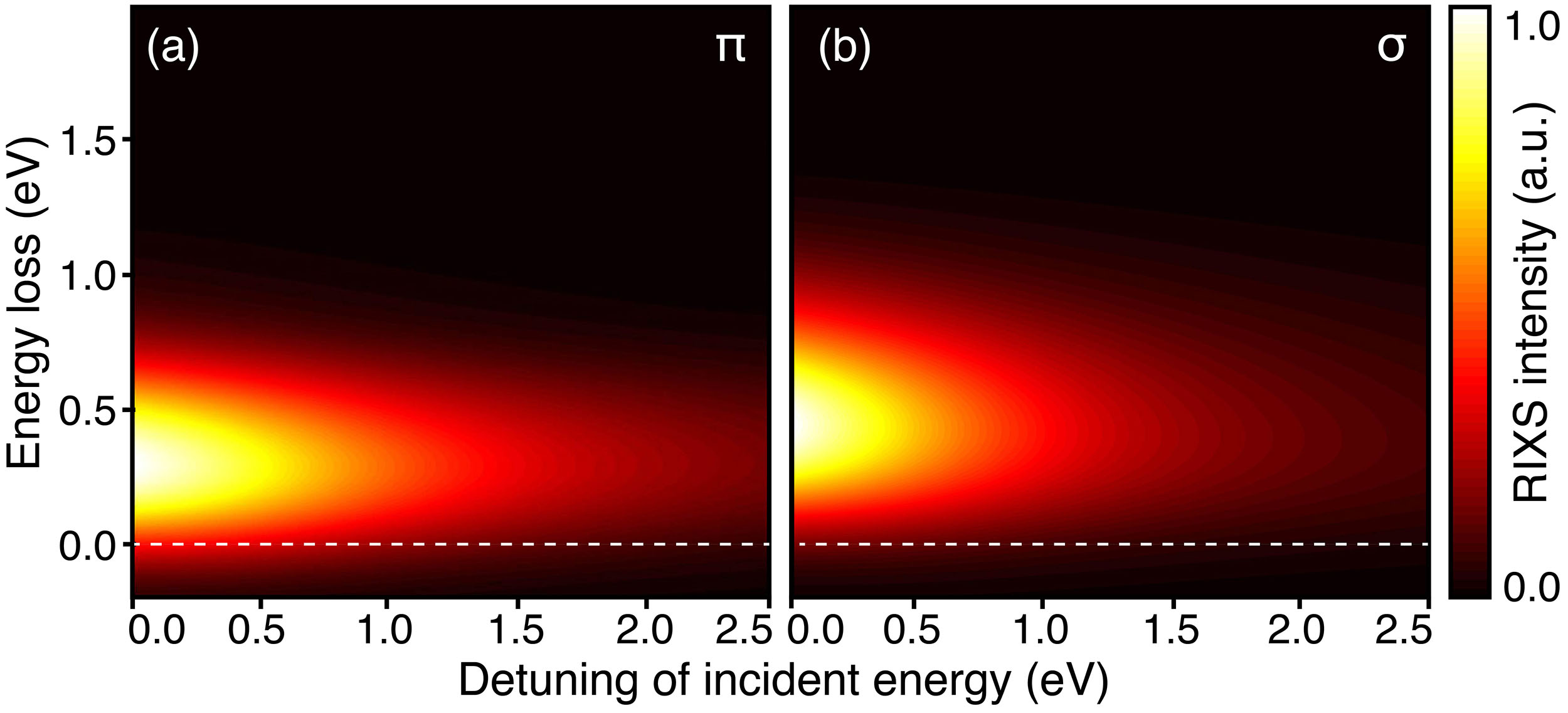}
\caption{(Color online.) 
Theoretical RIXS plots of the ${\rm YBCO+Ca}$ sample in terms of incoming photon energy
and energy loss with $\pi$ (a) and $\sigma$ (b) incoming polarization. 
[Parameters and scattering geometry are identical to the ones in Fig.~\ref{fig:SF_1D}~(f).]
}
\label{fig:2D_plots}
\end{figure}

Although we do find a minor shift in the peak positions in terms of detuning,  which is also visible in the experimental data, the shifts are rather insiginficant as compared to those found in Bi-2212.~\cite{Benjamin,Ronnow_vdBrink} These shifts come from the subtle interplay between the incoming photon energy $\omega$ and the energy loss $\Delta\omega$, and the underlying mechanism is most easily seen by considering the core hole free case ($U_0 = 0$).~\mprbnew{\cite{Carlisle}} Neglecting elastic contributions, the RIXS intensity can be written in this case as
\beq
I_{0} \, \propto \, \sum_{\bf k} \frac{n_F(\xi_{\bf k}) (1-n_F(\xi_{\bf k + q}))}{(\xi_{\bf k+q} - \omega)^2 + \Gamma^2}\;
\delta(\xi_{\bf k+q} - \xi_{\bf k} - \Delta\omega),
\label{eq:core_hole_free_RIXS}
\eeq
for both the SF and NSF channels.~\cite{Benjamin} Here $n_F$ denotes the Fermi function, and $\xi_{\bf k}$ stands for the quasi-particle energies measured from the Fermi surface. 
The above simple form suggests that the significant contributions to RIXS come from dynamically nested regions of filled electron states with empty states, shifted by a momentum $\bf q$ and an energy $\Delta \omega$.~\mprbnew{\cite{Carlisle, Carlisle2, Veenendaal,Jia, Denlinger}} Varying the incoming photon energy, changes the phase space available for the excitations, which modifies the contributions of different nesting regions, leading to shifts of the RIXS peaks~\cite{Benjamin} (see Appendix~\ref{app:dynamical_nesting}). This effect is incorporated in the denominator of Eq.~\eqref{eq:core_hole_free_RIXS}. Since the typical peaks reside at $300-500\, {\rm meV}$ energy, the RIXS response sensitively depends on the band structure in this energy range.~\mprbnew{\cite{Carlisle, Carlisle2, Shirley, Veenendaal}} The quasi-particle spectrum of YBCO and that of Bi-2212 in Ref.~\onlinecite{Benjamin} differ significantly here (see Appendix~\ref{app:compare_to_Bi2212}), making the peak shifts due to detuning predicted by Ref.~\onlinecite{Benjamin} for Bi-2212 larger than the ones observed by Minola {\it et al.}~\cite{Minola} for YBCO.

For further comparison, we plotted the RIXS intensities of the ${\rm YBCO + Ca}$ sample over a large range of detunings in Fig.~\ref{fig:2D_plots}, for both $\sigma$ ($97 \, \%$ NSF) and $\pi$ ($68\, \%$ SF) incoming polarizations. Taken into account the quasi-elastic peak and the high-energy tail of $dd$ transitions, our results agree well with experimental data in Figs.~4~(c,d) of Ref.~\onlinecite{Minola}.  We find that the peaks move only mildly with $\omega$ in both channels, whereas their intensities  diminish towards higher detunings, as expected. Although the experimentally observed peak in the $\sigma$ ($\sim$NSF) channel has been claimed to shift significantly with $\omega$, our simulations suggest that this effect is rather due to the superposition of the RIXS peak with the strong $dd$ background in the experimental data in Ref.~\onlinecite{Minola}.

\begin{table}[ht]
\caption{ARPES-based tight binding models of YBCO from Refs.~\onlinecite{BS_Norman, BS12, BS3}.}
\centering
\begin{tabular}{c c c c} 
\hline\hline
Hopping (meV) \hspace{10pt} & \hspace{10pt}BS1~\cite{BS3} \hspace{10pt} & \hspace{10pt} BS2~\cite{BS12} \hspace{10pt} & \hspace{10pt} BS3~\cite{BS_Norman} \hspace{10pt} \\ [0.5ex] 
\hline 
$t_{10}$ & -105 & -274 & -558 \\ 
$t_{11}$ &  29 & 140 &  273\\
$t_{20}$ & -25 & -19 &  -137\\
$t_{21}$ &  4  &  -13  &  \\
$t_{22}$ &   & 17 & \\ [1ex]
\hline
\end{tabular}
\label{table:BS}
\end{table}

\begin{figure}[t]
\includegraphics[width=8.5cm,clip=true]{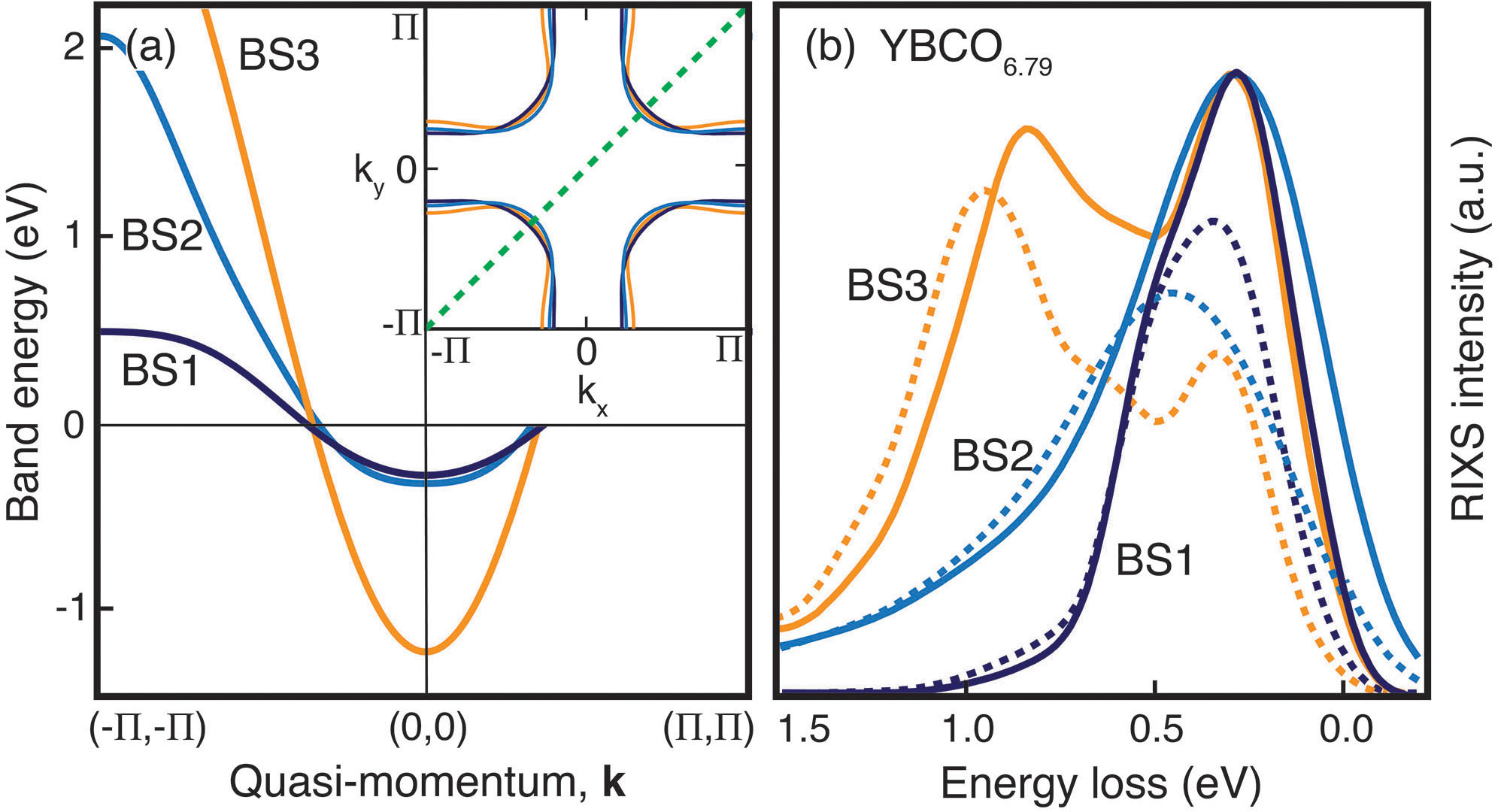}
\caption{(Color online.) 
Comparison of YBCO band structures, shown in Table.~\ref{table:BS}.
(a) Whereas the models agree well near the Fermi surface (inset), they are significantly different at higher energies. This leads to markedly different RIXS responses (b), with full (dashed) lines denoting theoretical RIXS spectra at incoming photon energy  0 (250) meV detuned from the XAS maximum.
[Parameters and scattering geometry are identical to the ones in Fig.~\ref{fig:polarization_resolved}~(a). Lattice size: $22 \times 22$ for BS1 and BS2, and $35 \times 35$ for BS3.]
}
\label{fig:BS}
\end{figure}

The sensitivity of RIXS on dynamical nesting makes it an unparalleled probe of high energy band structure. 
We use this opportunity to distinguish between different tight binding models of YBCO, shown in Table~\ref{table:BS}, by comparing their RIXS responses to the experimental data. \mprbnew{In contrast to ARPES, which is a direct probe of the band structure, RIXS measurements need to be compared to theoretical calculations that incorporate the effects of the core hole on the Fermi sea dynamics, in order to test different band structure models.~\cite{Veenendaal, Carlisle2, Benjamin}}
The band structure used in earlier figures is denoted by BS1. These models were obtained from fits to ARPES measurements near the Fermi energy, and show similar dispersions in this energy range. In contrast, due to the insensitivity of ARPES to higher lying excitations, they exhibit almost an order of magnitude differences near the top of the band, as shown in Fig.~\ref{fig:BS}~(a). As a result, their RIXS responses, shown in Fig.~\ref{fig:BS}~(b), are also significantly different. Whereas the band structures BS1 and BS2 produce similar results at resonant incoming energies, their peak widths and energy shifts are rather different at non-zero detunings. Band structure BS3 on the other hand produces additional peaks at higher energies, that are incompatible with the experimental data. Comparing our simulations to the experiment at all doping levels and at all detunings, we found, that band structure BS1 \mprb{of Ref.~\onlinecite{BS3}} \mprbnew{agrees most accurately with the experimental data} (see Fig.~\ref{fig:SF_1D} and Appendix~\ref{app:BS_comparison}). Thus, most likely, this band structure provides the most accurate picture of high energy electronic excitations of YBCO, among the models listed in Table~\ref{table:BS}.

In conclusion, we studied the recent RIXS experimental results of Minola {\it et al.}, Ref.~\onlinecite{Minola} in a quasi-particle theory,~\cite{Benjamin} over a wide range of dopings and detunings of the incoming photon energy, and found good agreement between theory and experiment.
We could thus explain the observed experimental features, originally attributed to  collective magnetic excitations, in terms of band structure theory.
We showed a natural physical picture of how the changes in incoming photon energy  leads to shifts in the RIXS peaks, due to the changing phase space of dynamical nesting regions, several hundred meV above the Fermi surface.
This makes RIXS a sensitive and versatile probe of the high energy excitations, as we demonstrated by comparing RIXS responses of three ARPES-based tight binding models of YBCO.
Importantly, the high energy part of the band structure had been inaccessible to traditional band structure measurement methods, such as ARPES. \mprbnew{Whereas ARPES reproduces the low energy band structure very accurately, combining it with momentum and incoming energy resolved RIXS measurements and calculations should provide a much more accurate description of the overall band structure.~\cite{RIXS_review, Carlisle, Veenendaal, Shirley, Carlisle2, Ahn}}
Given the technical advantages of RIXS, such as its insensitivity to surface effects, and its ability to probe sub-millimeter crystals and even films, it provides an unique opportunity to extend our knowledge of the electronic structure of high-temperature superconductors as well as other materials.\mprb{\cite{Jia, Denlinger, Kokko}}

\begin{acknowledgments}
\emph{Acknowledgments:}
Enlightening discussions with D. Benjamin, M. Le Tacon, B. Keimer, M. Norman, D. Chowdhury, P. Abbamonte, 
A. Keren, and D. Ellis are gratefully acknowledged.
The work of E. D. and M. K.-N. was supported by the Harvard-MIT CUA, NSF Grant No. DMR-1308435, 
AFOSR Quantum Simulation MURI, the ARO-MURI on Atomtronics, and ARO MURI Quism program.
E. D. also acknowledges support from Dr.~Max R\"ossler, the Walter Haefner Foundation and the ETH Foundation.
I. K. was supported by the NSF CAREER grant DMR-0956053.
\vfill \null
\end{acknowledgments}

\newpage

\appendix
\renewcommand{\theequation}{A\arabic{equation}}
\setcounter{equation}{0}  
\renewcommand{\thefigure}{A\arabic{figure}}
\setcounter{figure}{0}  

\section{Breakdown of quasi-particle theory in the antiferromagnetic phase} \label{app:AF_phase}

\begin{figure}
\includegraphics[width=5.5cm,clip=true]{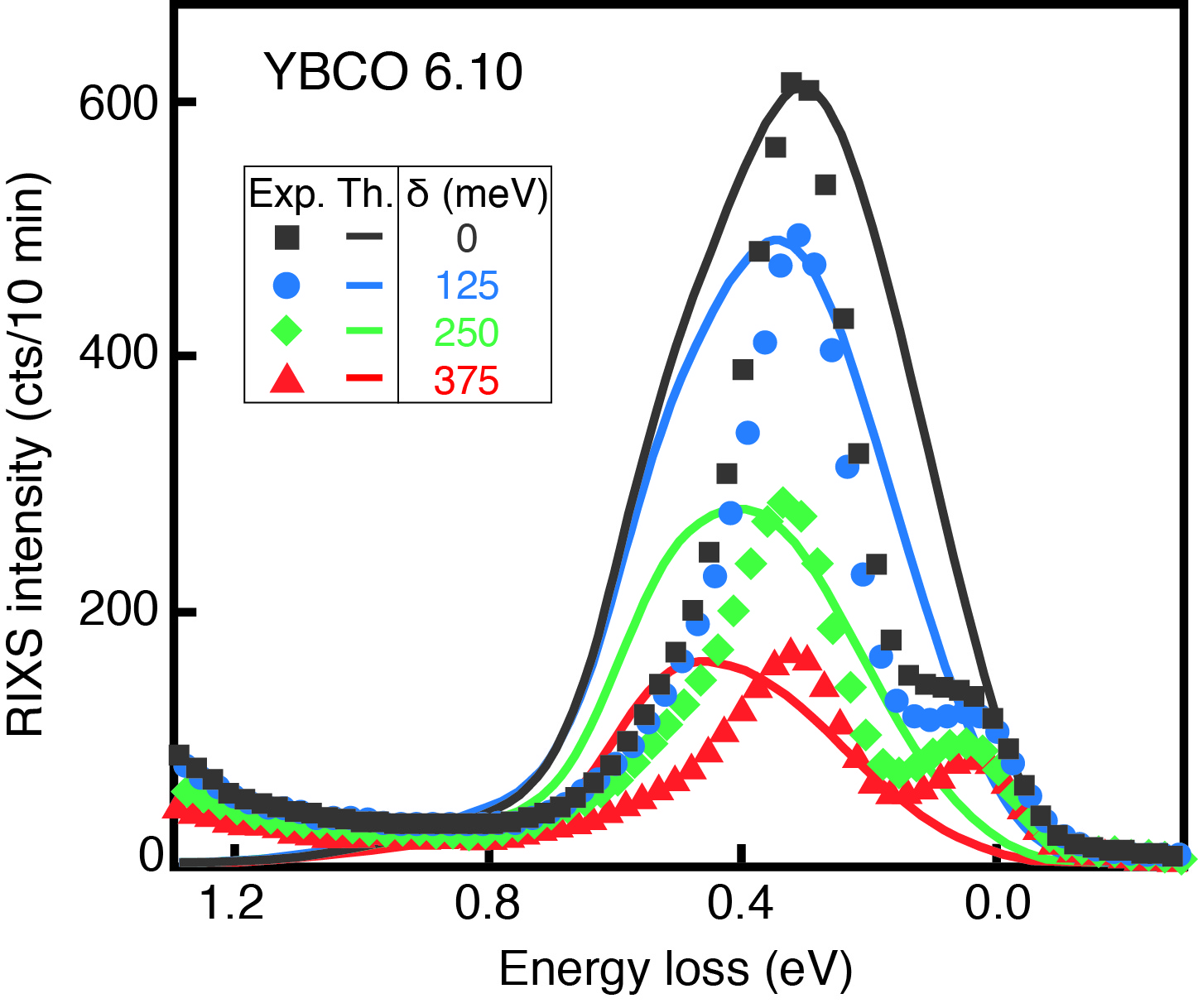}
\caption{(Color online.) RIXS spectra of the antiferromagnetic sample ${\rm YBCO}_{6.10}$
at $\pi$ incoming polarization, and at detunings $\delta = (0, 125, 250, 375) \, {\rm meV}$.
The experimental data shows pronounced peaks near $300 \, {\rm meV}$, together with elastic peaks
near zero energy.
The positions of the experimental peaks do not change with detuning, in contrast to the theoretical curves,
demonstrating that the quasi-particle theory cannot be valid in the antiferromagnetic phase.
[Parameters and scattering geometry are identical to the ones in Fig.~\ref{fig:polarization_resolved}~(a).]}
\label{fig:Supp_YBCO6_10}
\end{figure}

As we noted in the main text, the quasi-particle theory is expected 
to work best in the overdoped regime, and 
to break down near the antiferromagnetic phase, in the absence
of electronic quasi-particles. This is demonstrated in Fig.~\ref{fig:Supp_YBCO6_10},
where we compare theory with experiment on the almost undoped antiferromagnetic
sample, ${\rm YBCO}_{6.10}$. The experimental data exhibits significantly narrower
peaks than predicted by theory. Moreover, the centers of the peaks stay fixed 
near $300 \, {\rm meV}$, as incoming photon energy is detuned from the XAS maximum, 
in contrast to the theoretical predictions.

\section{Dynamical nesting} \label{app:dynamical_nesting}

\begin{figure}
\includegraphics[width=7.5cm,clip=true]{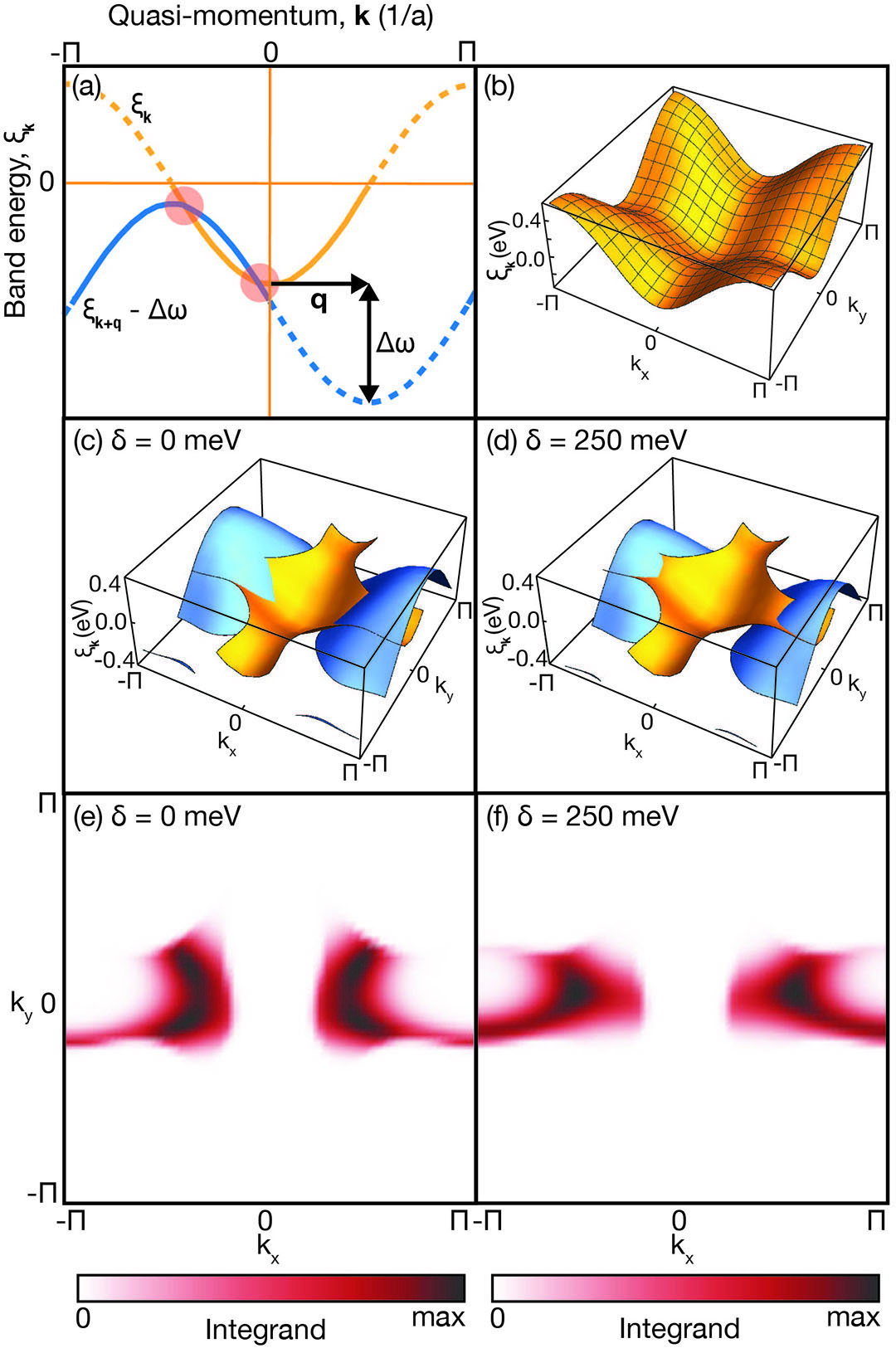}
\caption{(Color online.) 
Dynamical nesting regions contributing to the RIXS signal in case of zero core hole potential.
(a) Important contributions to RIXS come from regions where filled electron states (solid orange line)
intersect with empty states, shifted by an energy $\Delta \omega$ and momentum ${\bf q}$ (solid blue line).
(c, d) Dynamical nesting regions of the band structure BS1 (shown in (b)), at an incoming photon energy $\delta = 0$ and 250 $\rm meV$
detuned from the XAS maximum for (c) and (d) respectively. $\Delta \omega$ is chosen at maximum RIXS intensity.
(e, f) Corresponding integrand of Eq.~\eqref{eq:core_hole_free_RIXS} in $k$-space.
[Parameters and scattering geometry in (c-f) are identical to the ones in Fig.~\ref{fig:polarization_resolved}~(a),
except for the core hole potential $U_0 = 0~{\rm eV}$.]}
\label{fig:Supp_nesting}
\end{figure}

In order to understand how the interplay between incoming photon energy $\omega$
and energy loss $\Delta \omega$ affects the RIXS spectrum, it is worthwhile to consider the case of zero core hole 
potential, $U_0 = 0$. In this case, the RIXS intensity is given by Eq.~\eqref{eq:core_hole_free_RIXS}.
As denoted schematically in Fig.~\ref{fig:Supp_nesting}~(a), the important contributions to the RIXS 
signal come from dynamical nesting regions, where occupied electronic states (solid orange line)
intersect with empty states shifted by a momentum ${\bf q}$ and an energy $\Delta \omega$ (solid blue line).
The nesting regions are shown in Figs.~\ref{fig:Supp_nesting}~(c-d) in case of the band structure BS1.
As $\omega$ is tuned, the phase space of these excitations changes, and the nesting 
regions get different weights through
the denominator of Eq.~\eqref{eq:core_hole_free_RIXS}. This is demonstrated in Figs.~\ref{fig:Supp_nesting}~(e-f),
showing the integrand of Eq.~\eqref{eq:core_hole_free_RIXS} in $k$-space.

\section{Comparison with Bi-2212} \label{app:compare_to_Bi2212}

\begin{figure}
\includegraphics[width=7.5cm,clip=true]{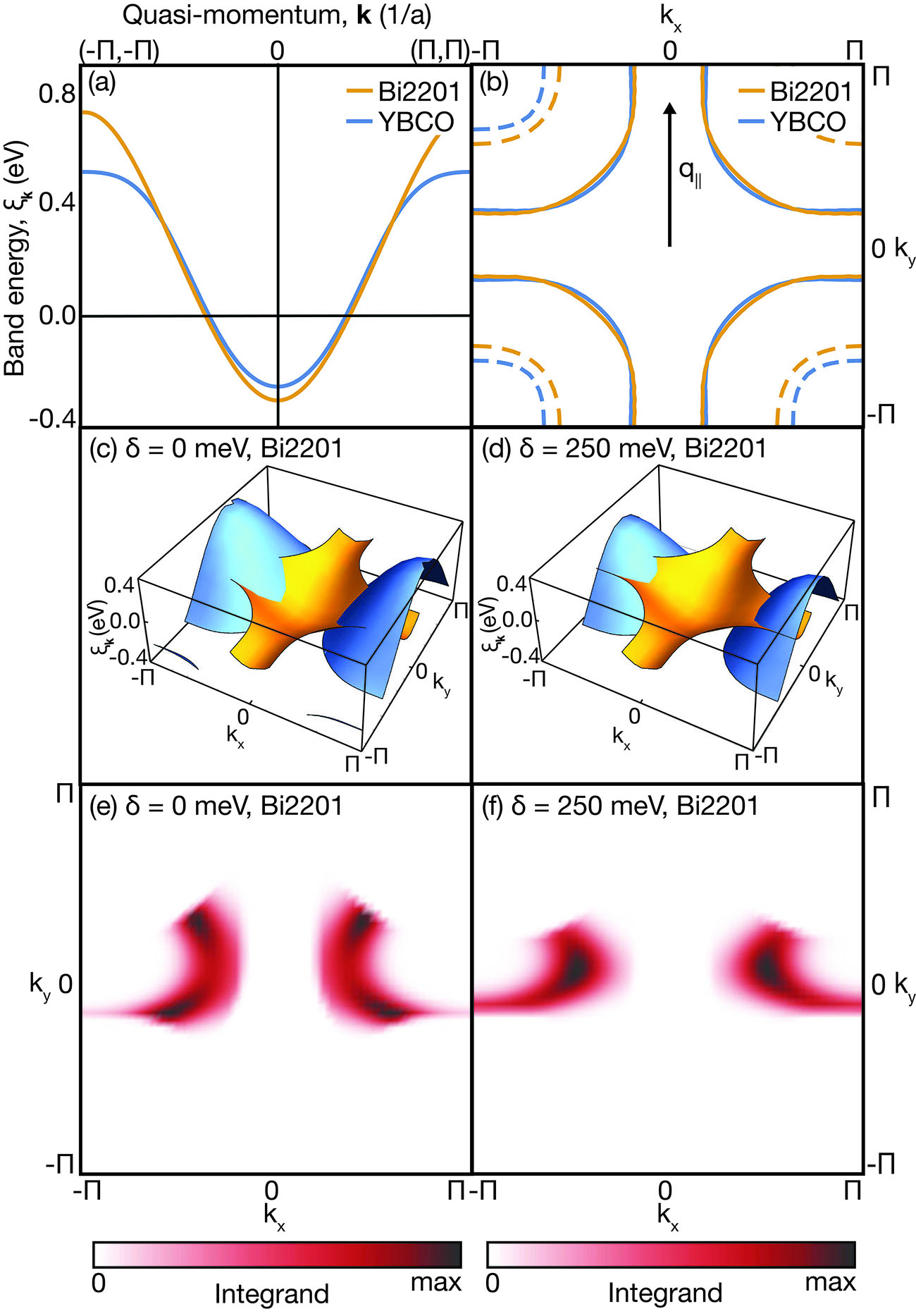}
\caption{
(Color online.) 
(a) Comparison of the band energies BS1 for YBCO (blue) and that of Bi-2212 (orange)~[18] along the ${\bf k} \parallel (1,1)$ direction.
(b) Fermi surface (solid lines) and and equienergetic surfaces $450 \, {\rm meV}$ above the Fermi level (dashed lines), with orange (blue) corresponding to Bi-2212 (YBCO).
(c, d) Dynamical nesting regions for Bi-2212 at an incoming photon energy $\delta = 0$ and
250 meV detuned from the XAS maximum for (c) and (d) re-
spectively. $\Delta\omega$ is chosen at maximum RIXS intensity. (e, f)
Corresponding integrand of Eq.~\eqref{eq:core_hole_free_RIXS} in $k$-space.
[Parameters: ${\bf q}_\parallel = 2\pi \,(0, 0.37)$ as shown in (b), $U_0 = 0 \, {\rm eV}$, 
$\Gamma = 250 \, {\rm meV}$, energy resolution 95 meV HWHM.]
}
\label{fig:Supp_Bi2212}
\end{figure}

Fig.~\ref{fig:Supp_Bi2212}~(a,b) compare the band structure of Bi-2212 used by Ref.~\onlinecite{Benjamin} 
to BS1 of YBCO (see Table~I). Whereas the two band structures are very similar near the Fermi surface,
they differ significantly near the top of the band at $400-500~\rm{meV}$. 
As we pointed out in the main text, since the RIXS peaks reside typically in this energy range, 
dynamical nesting at these energies is largely responsible for the shape and position of the RIXS peak. 
Fig.~\ref{fig:Supp_nesting}~(c-f) and Fig.~\ref{fig:Supp_Bi2212}~(c-f) exhibit the dynamical nesting regions 
and the momentum space contributions to the RIXS intensity at different detunings in case of YBCO and Bi-2212, respectively, 
showing apparent differences in the shapes of the nesting regions. 
Since the top of the band for the Bi-2212 band structure is higher, phase space is available for excitations of higher energy 
than those of the typical RIXS process in YBCO. This generally leads to broader peaks and more significant shifts in peak position
as the incoming energy is detuned, in accordance with the findings of~Ref.~\onlinecite{Benjamin}.

\section{Comparison of YBCO band structures} \label{app:BS_comparison}

\begin{figure}
\includegraphics[width=8.5cm,clip=true]{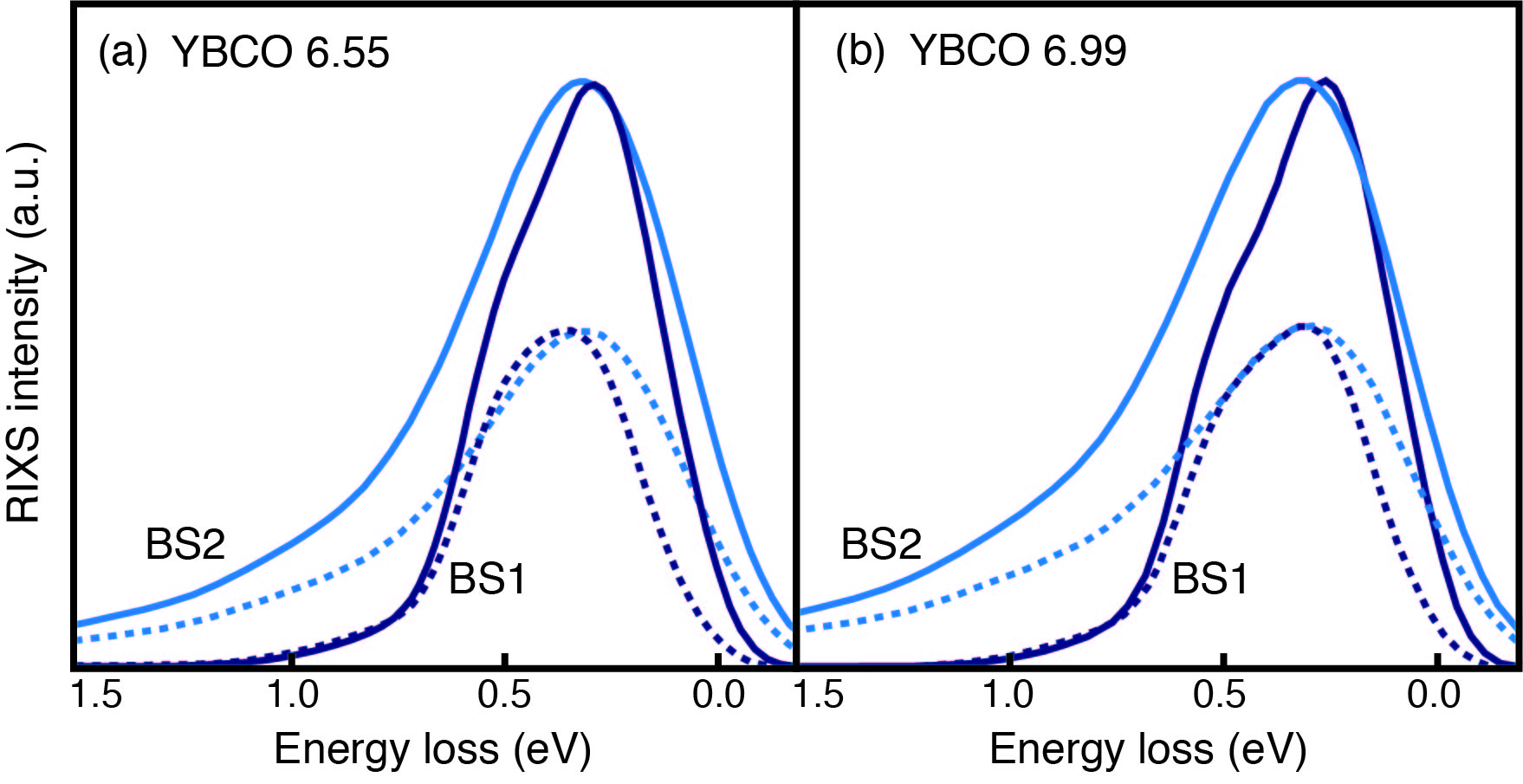}
\caption{(Color online.) Comparison of band structures BS1 and BS2 
in case of underdoped ${\rm YBCO}_{6.55}$ and the optimally doped ${\rm YBCO}_{6.99}$.
Full (dashed) lines show RIXS spectra at $\delta = 0 \, (250) \, {\rm meV}$ detunings.
[Parameters and scattering geometry are identical to the ones in Fig.~\ref{fig:polarization_resolved}~(a).]}
\label{fig:Supp_BS12}
\end{figure}

In Fig.~\ref{fig:Supp_BS12}, we present further theoretical RIXS curves to compare band structures BS1 and BS2,
shown in Table~I. 
The peaks have been rescaled by an overall factor for comparison.
Both for underdoped ${\rm YBCO}_{6.55}$ and the optimally doped ${\rm YBCO}_{6.99}$, 
BS1 produces narrower peaks, that are also at somewhat
lower energies at zero detuning. This gives a better fit to experimental data, shown in Fig.~\ref{fig:SF_1D}.


\begin{thebibliography}{}
\bibitem{LDA_calculations}  S. Massidda, J. Yu, A. J. Freeman, and D. D. Koelling, Phys. Lett. A, {\bf 122}, 198 (1987).
\bibitem{ARPES_review} A. Damascelli, Z. Hussain, and Z.-X. Shen, Rev. Mod. Phys. {\bf 75}, 473 (2003).
\bibitem{Q_oscillation_1} B. Vignolle, A. Carrington, R. A. Cooper, M. M. J. French, A. P. Mackenzie, C. Jaudet, D. Vignolles, Cyril Proust and N. E. Hussey, Nature {\bf 455}, 952 (2008).
\bibitem{Q_oscillation_2} N. Doiron-Leyraud, C. Proust, D. LeBoeuf, J Levallois, J.-B. Bonnemaison, R. Liang, D. A. Bonn, W. N. Hardy and L. Taillefer, Nature {\bf 447}, 565 (2012).
\bibitem{BS_Norman} Y.~Sassa, M.~Radovic, M.~Mansson, E.~Razzoli, X.~Y.~Cui, S.~Pailhes, S.~Guerrero, M.~Shi, P.~R.~Willmott, F.~Miletto~Granozio, J.~Mesot, M.~R.~Norman, L.~Patthey, Phys.~Rev.~B {\bf 83}, 140511(R) (2011).
\bibitem{BS3} K. Pasanai and W. A. Atkinson, Phys. Rev. B {\bf 81,} 134501 (2010).
\bibitem{BS12} M. C. Schabel, C.-H. Park, A. Matsuura, Z.-X. Shen, D. A. Bonn, R. Liang and W. N. Hardy, Phys. Rev. B {\bf 57}, 6090 (1998).
\bibitem{electronic_RIXS} M. Le Tacon, A. Bosak, S. M. Souliou, G. Dellea, T. Loew, R. Heid, K-P. Bohnen, G. Ghiringhelli, M. Krisch and B. Keimer, Nat. Phys. {\bf 10}, 52 (2014).
\bibitem{monney} C.~Monney, K.~J.~Zhou, H.~Cercellier, Z.~Vydrova, M.~G.~Garnier, G.~Monney, V.~N.~Strocov, H.~Berger, H.~Beck, T.~Schmitt, and P.~Aebi, Phys. Rev. Lett. {\bf 109}, 047401 (2012).
\bibitem{magnetic_RIXS} Y. Y. Peng, M. Hashimoto, M. Moretti Sala, A. Amorese, N. B. Brookes, G. Dellea, W.-S. Lee, M. Minola, T. Schmitt, Y. Yoshida, K.-J. Zhou, H. Eisaki, T. P. Devereaux, Z.-X. Shen, L. Braicovich, G. Ghiringhelli, arXiv:1504.05165.
\bibitem{phononic_RIXS} W.~S.~Lee, S.~Johnston, B.~Moritz, J.~Lee, M.~Yi, K.~J.~Zhou, T.~Schmitt, L.~Patthey, V.~Strocov, K.~Kudo, Y.~Koike, J.~van~den~Brink, T.~P.~Devereaux, and Z.~X.~Shen, Phys. Rev. Lett. {\bf 110}, 265502 (2013).
\bibitem{RIXS_film} M. P. M. Dean,	R. S. Springell,	C. Monney,	K. J. Zhou,	J. Pereiro,	I. Bozovic, B. Dalla Piazza, H. M. Ronnow,	 E. Morenzoni,	J. van den Brink,	T. Schmitt and J. P. Hill, Nat. Mat. {\bf 11}, 850 (2012).
\bibitem{RIXS_review} A. Kotani and S. Shin, Rev. Mod. Phys. {\bf 73}, 203 (2001).
\bibitem{Carlisle} \mprb{ J. A. Carlisle, Eric L. Shirley, E. A. Hudson, L. J. Terminello, T. A. Callcott, J. J. Jia, D. L. Ederer, R. C. C. Perera, and F. J. Himpsel, Phys. Rev. Lett. {\bf 74}, 1234 (1995).}
\bibitem{Carlisle2} \mprbnew{J. A. Carlisle, Eric L. Shirley, L. J. Terminello, J. J. Jia, T. A. Callcott, D. L. Ederer, R. C. C. Perera, and F. J. Himpsel
Phys. Rev. B {\bf 59}, 7433 (1999).}
\bibitem{Veenendaal} \mprb{ M. van Veenendaal and P. Carra, Phys. Rev. Lett. {\bf 78}, 2839 (1997).}
\bibitem{Shirley} \mprb{ E. L. Shirley, Phys. Rev. Lett. {\bf 80}, 794 (1998).}
\bibitem{Ahn} \mprb{ K. H. Ahn, A. J. Fedro, and M. van Veenendaal, Phys. Rev. B {\bf 79}, 045103 (2009).}
\bibitem{electronicQPs} M. R. Norman, H. Ding, M. Randeria, J. C. Campuzano, T. Yokoya, T. Takeuchi, T. Takahashi, T. Mochiku, K. Kadowaki, P. Guptasarma and D. G. Hinks, Nature {\bf 392}, 157 (1998).
\bibitem{Fermi_surface_HTC1} D. LeBoeuf, N. Doiron-Leyraud, J. Levallois, R. Daou, J.-B. Bonnemaison, N. E. Hussey, L. Balicas, B. J. Ramshaw, Ruixing Liang, D. A. Bonn, W. N. Hardy, S. Adachi, Cyril Proust and Louis Taillefer, Nature {\bf 450}, 533 (2007).
\bibitem{Fermi_surface_HTC2} N. Doiron-Leyraud, C. Proust, D. LeBoeuf, J. Levallois, J.-B. Bonnemaison, R. Liang, D. A. Bonn, W. N. Hardy and L. Taillefer, Nature {\bf 447}, 565 (2007).
\bibitem{Fermi_surface_HTC3}  B. Vignolle, D. Vignolles, D. LeBoeuf, S. Lepault, B. Ramshaw,
R. Liang, D. A. Bonn, W.N. Hardy, N. Doiron-Leyraud, A. Carrington, N. E. Hussey, L. Taillefer, C. Proust, C. R. Phys. {\bf 12}, 446 (2011).
\bibitem{Benjamin} D. Benjamin, I. Klich and E. Demler, Phys. Rev. Lett. {\bf 112}, 247002 (2014).
\bibitem{spin_susc_RIXS_calculations0} M. W. Haverkort, Phys. Rev. Lett. {\bf 105}, 167404 (2010).
\bibitem{spin_susc_RIXS_calculations1} L. J. P. Ament, M. van Veenendaal, T. P. Devereaux, J.
P. Hill, and J. van den Brink, Rev. Mod. Phys. {\bf 83}, 705 (2011).
\bibitem{spin_susc_RIXS_calculations2} L. J. P. Ament, G. Ghiringhelli, M. M. Sala, 
L. Braicovich, and J. van den Brink, Phys. Rev. Lett. {\bf 103}, 117003 (2009).
\bibitem{spin_susc_RIXS_calculations3} M. W. Haverkort, Phys. Rev. Lett. {\bf 105}, 167404 (2010).
\bibitem{spin_susc_RIXS_calculations4} C. J. Jia, E. A. Nowadnick, K. Wohlfeld, Y. F. Kung, 
C. C. Chen, S. Johnston, T. Tohyama, B. Moritz, and T. P. Devereaux, Nat. Commun. {\bf 5}, 3314 (2014).
\bibitem{sr2cuo3_spin_charge} J. Schlappa,	K. Wohlfeld,	K. J. Zhou,	M. Mourigal,	M. W. Haverkort,	V. N. Strocov,	L. Hozoi,	C. Monney,	S. Nishimoto,	S. Singh,	A. Revcolevschi,	J.-S. Caux,	L. Patthey,	H. M. Ronnow,	J. van den Brink and T. Schmitt, Nature {\bf 485}, 82 (2012).
\bibitem{D_Benjamin_exp1} M. Le Tacon, M. Minola, D. C. Peets, M. Moretti Sala, S. Blanco-Canosa, V. Hinkov, R. Liang, D. A. Bonn, W. N. Hardy, C. T. Lin, T. Schmitt, L. Braicovich, G. Ghiringhelli, and B. Keimer, Phys. Rev. B {\bf 88}, 020501(R) (2013).
\bibitem{D_Benjamin_exp2} M. P. M. Dean, A. J. A. James, R. S. Springell, X. Liu, C. Monney, K. J. Zhou, R. M. Konik, J. S. Wen, Z. J. Xu, G. D. Gu, V. N. Strocov, T. Schmitt, and J. P. Hill, Phys. Rev. Lett. {\bf 110}, 147001 (2013).
\bibitem{Anderson} P. W. Anderson, Phys. Rev. Lett. {\bf 18}, 1049 (1967).
\bibitem{Nozieres} P. Nozieres and E. Abrahams, Phys. Rev. B {\bf 10}, 3099 (1974).
\bibitem{Ronnow_vdBrink} M. Guarise, B. Dalla Piazza,	H. Berger,	E. Giannini,	T. Schmitt,	H. M. Ronnow,	G. A. Sawatzky,	J. van den Brink,	D. Altenfeld,	 I. Eremin and M. Grioni, Nat. Commun. {\bf 5}, 5760 (2014).
\bibitem{Abbamonte} P. Abbamonte, C. A. Burns, E. D. Isaacs, P. M. Platzman, L. L. Miller, S. W. Cheong, and M. V. Klein, Phys. Rev. Lett.~{\bf 83}, 860 (1999).
\bibitem{Minola} M. Minola, G. Dellea, H. Gretarsson, Y. Y. Peng, Y. Lu, J. Porras, T. Loew, F. Yakhou, N. B. Brookes, Y. B. Huang, J. Pelliciari, T. Schmitt, G. Ghiringhelli, B. Keimer, L. Braicovich, and M. Le Tacon, Phys. Rev. Lett. {\bf 114}, 217003 (2015).
\bibitem{QPlifetime_footnote} The theory based on quasi-particles is expected to work as long as their width 
is significantly smaller than $\Gamma$, which holds even for energies well above the Fermi surface.~\cite{BS12,QPlifetime}
\bibitem{OC_physics1} Gerald D. Mahan, Many Particle Physics (Kluwer, New York, 2000), 3rd ed.
\bibitem{OC_physics2} Yuli V. Nazarov and Yaroslav M. Blanter, Quantum Transport: Introduction to Nanoscience (Cambridge
University Press, Cambridge, 2009), 1st ed.
\bibitem{Klich_formulas} I. Klich, in {\it Quantum Noise in Mesoscopic Physics}, edited
by Y. Nazarov (Springer, New York, 2003).
\bibitem{Devereaux} \mprb{ H. Y. Huang, C. J. Jia, Z. Y. Chen, K. Wohlfeld, B. Moritz, T. P. Devereaux, W. B. Wu, J. Okamoto, W. S. Lee, M. Hashimoto, Y. He, Z. X. Shen, Y. Yoshida, H. Eisaki, C. Y. Mou, C. T. Chen and D. J. Huang, Sci. Rep. {\bf 6}, 19657 (2016).}
\bibitem{Keimer_more_friendly} \mprb{ C. Monney, T. Schmitt, C. E. Matt, J. Mesot, V. N. Strocov, O. J. Lipscombe, S. M. Hayden, and J. Chang, Phys. Rev. B {\bf 93}, 075103 (2016).}
\bibitem{Jia} \mprb{ J. J. Jia, T. A. Callcott, Eric L. Shirley, J. A. Carlisle, L. J. Terminello, A. Asfaw, D. L. Ederer, F. J. Himpsel, and R. C. C. Perera, Phys. Rev. Lett. {\bf 76}, 4054 (1996).}
\bibitem{Denlinger} \mprb{ J. D. Denlinger, J. A. Clack, J. W. Allen, G.-H. Gweon, D. M. Poirier, C. G. Olson, J. L. Sarrao, A. D. Bianchi, and Z. Fisk, Phys. Rev. Lett. {\bf 89}, 157601 (2002).}
\bibitem{Kokko} \mprb{ K. Kokko, V. Kulmala, J. A. Leiro, and W. Hergert, Phys. Rev. B {\bf 68}, 052503 (2003).}
\bibitem{QPlifetime} D. C. Peets, J. D. F. Mottershead, B. Wu, I. S. Elfimov, R. Liang, W. N. Hardy, D. A. Bonn, M. Raudsepp, N. J. C. Ingle and A. Damascelli, New. J. Phys. {\bf 9}, 28 (2007). 
\bibliography{apssamp}
\end{thebibliography}
\end{document}